\UseRawInputEncoding
\documentclass[aps,prl,twocolumn,superscriptaddress,showpacs]{revtex4-2}
\usepackage{graphicx}
\usepackage{dcolumn}
\usepackage{bm}
\usepackage{xcolor}

\newcommand{\ef}{$E_F$}

\begin{document}


\title{Non-Thermal Emergence of an Orbital-Selective Mott Phase in FeTe$_{1-x}$Se$_x$}

\author{Jianwei Huang}
\affiliation{Department of Physics and Astronomy, Rice Center for Quantum Materials, Rice University, Houston, Texas 77005, USA}
\author{Rong Yu}
\affiliation{Department of Physics, Renmin University of China, Beijing 100872, China}
\author{Zhijun Xu}
\affiliation{NIST Center for Neutron Research, National Institute of Standards and Technology, Gaithersburg Maryland 20899, USA}
\affiliation{Department of Materials Science and Engineering, University of Maryland, College Park, Maryland 20742, USA}
\affiliation{Department of Physics, University of California Berkeley, Berkeley, California 94720, USA}
\author{Jian-Xin Zhu}
\affiliation{Theoretical Division, Los Alamos National Laboratory, Los Alamos, New Mexico 87545, USA}
\affiliation{Center for Integrated Nanotechnologies, Los Alamos National Laboratory, Los Alamos, New Mexico 87545, USA}
\author{Qianni Jiang}
\affiliation{Department of Physics, University of Washington, Seattle, Washington 98195, USA}
\author{Meng Wang}
\affiliation{School of Physics, Sun Yat-sen University, Guangzhou, Guangdong 510275, China}
\author{Han Wu}
\affiliation{Department of Physics and Astronomy, Rice Center for Quantum Materials, Rice University, Houston, Texas 77005, USA}
\author{Tong Chen}
\affiliation{Department of Physics and Astronomy, Rice Center for Quantum Materials, Rice University, Houston, Texas 77005, USA}
\author{Jonathan D. Denlinger}
\affiliation{Advanced Light Source, Lawrence Berkeley National Lab, Berkeley, California 94720, USA}
\author{Sung-Kwan Mo}
\affiliation{Advanced Light Source, Lawrence Berkeley National Lab, Berkeley, California 94720, USA}
\author{Makoto Hashimoto}
\affiliation{Stanford Synchrotron Radiation Lightsource, SLAC National Accelerator Laboratory, Menlo Park, California 94025, USA}
\author{Genda Gu}
\affiliation{Condensed Matter Physics and Materials Science Department, Brookhaven National Laboratory, Upton, New York, USA}
\author{Pengcheng Dai}
\affiliation{Department of Physics and Astronomy, Rice Center for Quantum Materials, Rice University, Houston, Texas 77005, USA}
\author{Jiun-Haw Chu}
\affiliation{Department of Physics, University of Washington, Seattle, Washington 98195, USA}
\author{Donghui Lu}
\affiliation{Stanford Synchrotron Radiation Lightsource, SLAC National Accelerator Laboratory, Menlo Park, California 94025, USA}
\author{Qimiao Si}
\affiliation{Department of Physics and Astronomy, Rice Center for Quantum Materials, Rice University, Houston, Texas 77005, USA}
\author{Robert J. Birgeneau}
\email{robertjb@berkeley.edu}
\affiliation{Department of Physics, University of California Berkeley, Berkeley, California 94720, USA}
\affiliation{Materials Sciences Division, Lawrence Berkeley National Laboratory, Berkeley, California 94720, USA}
\affiliation{Department of Materials Science and Engineering, University of California, Berkeley, USA}
\author{Ming Yi}
\email{mingyi@rice.edu}
\affiliation{Department of Physics and Astronomy, Rice Center for Quantum Materials, Rice University, Houston, Texas 77005, USA}
\affiliation{Department of Physics, University of California Berkeley, Berkeley, California 94720, USA}

\date{\today}

\begin{abstract}
Electronic correlation is of fundamental importance to high temperature superconductivity. Iron-based superconductors are believed to possess moderate correlation strength, which combined with their multi-orbital nature makes them a fascinating platform for the emergence of exotic phenomena. A particularly striking form is the emergence of an orbital selective Mott phase, where the localization of a subset of orbitals leads to a drastically reconstructed Fermi surface. Here, we report spectroscopic evidence of the reorganization of the Fermi surface from FeSe to FeTe as Se is substituted by Te. We uncover a particularly transparent way to visualize the localization of the $d_{xy}$ electron orbital through the suppression of its hybridization with the more coherent $d$ electron orbitals, which leads to a redistribution of the orbital-dependent spectral weight near the Fermi level. These noteworthy features of the Fermi surface are accompanied by  a divergent behavior of a band renormalization in the $d_{xy}$ orbital. All of our observations are further supported by our theoretical calculations to be salient spectroscopic signatures of such a non-thermal evolution from a strongly correlated metallic phase towards an orbital-selective Mott phase in FeTe$_{1-x}$Se$_x$ as Se concentration is reduced. 

\end{abstract}

\maketitle


\section{INTRODUCTION}
In multi-orbital quantum materials, orbital-dependent correlations and interactions often lead to exotic emergent phenomena. An extreme case is in the heavy fermion systems where the strongly localized \textit{f} electrons coexist with itinerant electrons, the interactions between which give rise to an abundance of competing magnetic ground states and unconventional superconductivity in proximity to quantum criticality~\cite{Kirchner2020}. Orbital-dependent correlation effects have also been reported to play important roles in Ca$_{2-x}$Sr$_x$RuO$_4$~\cite{Anisimov2002}, VO$_2$~\cite{Mukherjee2016}, and transition metal dichalcogenide~\cite{Qiao2017}, primarily in a regime where selected orbitals experience significant mass enhancements towards a Mott-insulating limit while the remaining orbitals maintain a degree of itinerancy. Such a phase is known as an orbital-selective Mott phase (OSMP)~\cite{DeMedici2005,Ferrero2009,Yin2011,Greger2013,Yu2013a,Georges2013,DeMedici2014}. As the system approaches an OSMP through the tuning of a non-thermal control parameter at low temperatures, the low energy electronic states are expected to undergo dramatic reconstruction such as redistribution in momentum space, leading to signatures in the transport behaviors. Such an evolution, while in principle observable via momentum-resolved spectroscopy, has proven to be difficult to resolve due to either the associated small energy scales~\cite{Danzenbacher2007,Fujimori2012,Neupane2015,Chen2017,Leuenberger2018} or incompatibility of experimental implementation of tuning parameters~\cite{Gegenwart2008,Si2010,Mun2013}. Indeed, in the cases where the OSMP has been well established as a competing ground state, i.e. those of heavy fermion metals, there has never been an observation of the OSMP's reconstructed Fermi surface (FS) by angle-resolved photoemission spectroscopy (ARPES).

Recently, orbital-dependent correlation effects have been found in iron-based superconductors (FeSCs)~\cite{Yin2011,Si2016}. In particular, while the FeSCs as bad metals are deemed to be moderately correlated compared to the cuprate high temperature superconductors, the electrons belonging to different orbitals are found to be correlated to different degrees -- stronger in the $d_{xy}$ orbital than in the $d_{xz}$/$d_{yz}$ orbitals. This orbital-differentiation is enhanced by Hund’s coupling, and increases systematically in sync with the vertical elongation of the iron tetrahedron from the iron-phosphides to iron-pnictides to iron-chalcogenides~\cite{Yi2017a}, in which strong orbital-selectivity has been reported~\cite{Yin2012,Yi2015}. From the electronic degree of freedom, the effective mass of the $d_{xy}$ orbital dominated band has been reported to be larger than that of the $d_{xz}$/$d_{yz}$ orbitals. From both nuclear magnetic resonance and neutron scattering measurements, coexistence of both itinerant and local spins has been found~\cite{Dai2012}, where the $d_{xy}$ orbital contributes dominantly to the local spin susceptibility~\cite{Li2020,Li2018,Song2018}. For example, inelastic neutron scattering experiments on detwinned NaFeAs, a parent compound of FeSCs, have shown that spin waves of the system are orbital selective with high energy spin waves arise mostly from the $d_{xy}$ orbital and obey the C$_4$ rotational symmetry, while the low energy spin wave are from the $d_{xz}$/$d_{yz}$ orbitals that break the C$_4$ rotational symmetry below the nematic phase transition temperature~\cite{Tam2020}.  In addition, the neutron spin resonance coupled with superconductivity, seen in most FeSCs~\cite{Dai2015a}, seems to be mostly associated with the $d_{yz}$ orbitals, thus suggesting orbital selective superconductivity~\cite{Tian2019}. Previous works, both experimental and theoretical, have shown that when the material system exhibits sufficient orbital-selectivity in the ground state, it can crossover into an orbital-selective incoherent state or OSMP with increasing temperature~\cite{Hardy2013,Miao2016,Yang2017}. 

\begin{figure}
\includegraphics[width=0.48\textwidth]{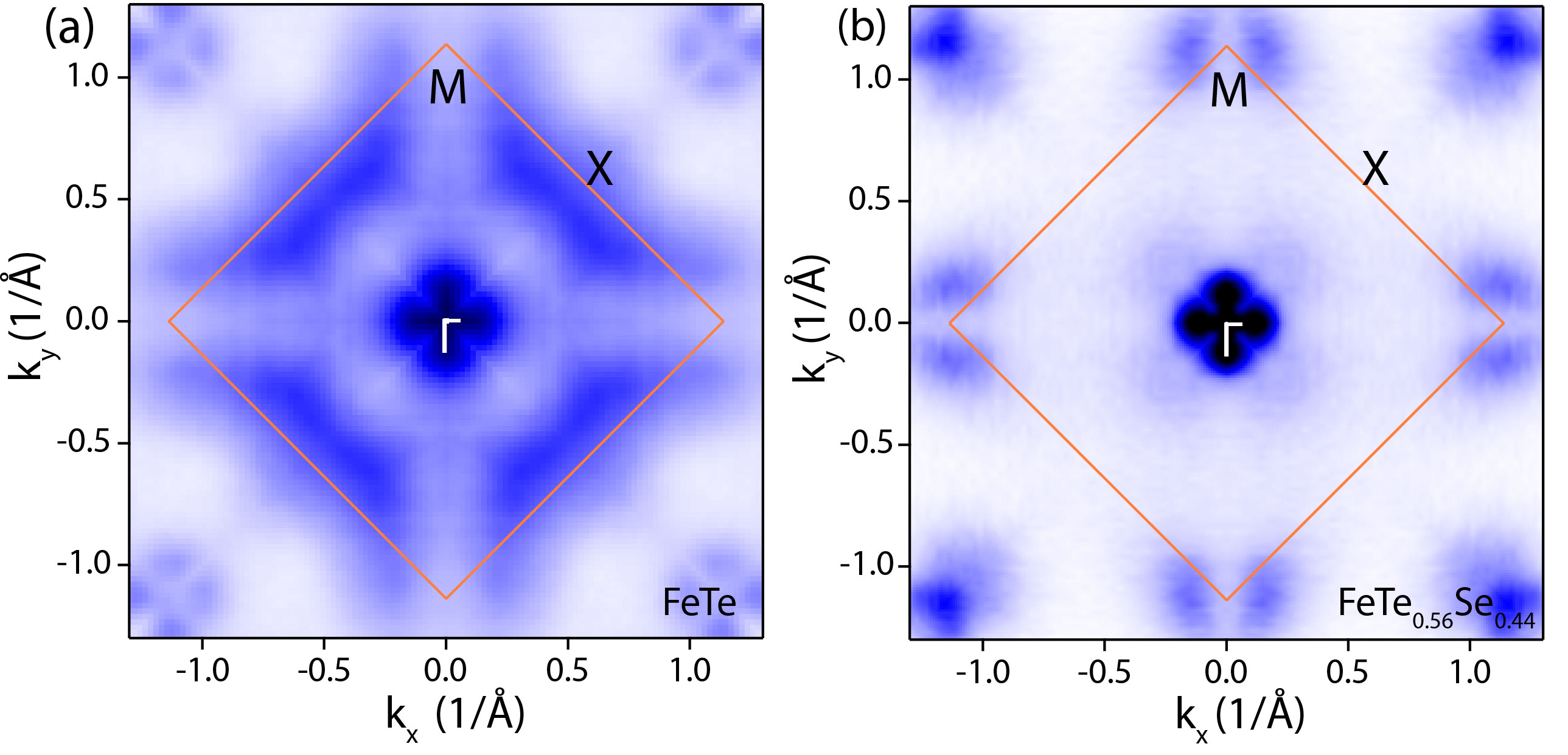}
\caption{\label{fig:Fig1}Fermi surfaces of FeTe and FeTe$_{0.56}$Se$_{0.44}$. (a) Measured Fermi surfaces for FeTe taken at 80K. (b) Measured FS for FeTe$_{0.56}$Se$_{0.44}$ taken at 10K. Both maps were measured with 72eV photons and symmetrized according to C$_4$ rotational symmetry of the crystal structure.}
\end{figure}

These developments set the stage to address a crucial question that remains open for the FeSCs: Is there an OSMP among the ground states that compete against the all-orbital-itinerant metallic ground state? To proceed, one must go beyond the study of any thermal crossover and, instead, scan the ground state landscape by tuning a non-thermal control parameter while staying at low temperatures. An added benefit of doing the latter is that the FS is only sharply defined at low temperatures, and studying the quantum phase transition through a non-thermal parameter variation allows for unambiguously detecting any FS reconstruction. More broadly, it is particularly opportune to address this issue in 3$d$-electron systems such as the FeSCs. These systems have energy scales that allow for a direct momentum-space visualization -- by ARPES -- of the Fermi surface and its reconstruction across their ground states, a feature that is not possible for alternative materials platforms with much smaller energy scales such as for the heavy fermion systems as mentioned earlier.

In this work, we present systematic evidence for a low-temperature tunability towards an OSMP in the FeTe$_{1-x}$Se$_x$ family of superconductors without raising temperatures. Remarkably, strong reorganization of the FS is observed in the absence of any symmetry-breaking electronic order. We demonstrate how the more coherent $d$-orbitals, in particular, the $d_{z^2}$ orbital, serve as a transparent means to visualize the suppression of the $d_{xy}$ states at the Fermi energy. In addition, we show that the FS reconstruction is accompanied by an orbital-dependent mass enhancement, and signatures of lower Hubbard band as we tune from the FeSe end towards FeTe. Combined with the temperature axis, we arrive at a comprehensive phase diagram of the orbital-selectivity in FeTe$_{1-x}$Se$_x$, the understanding of which is discussed in connection with anomalies reported in measurements of the Hall coefficient, resistivity, and magnetic excitations.

\begin{figure*}
\includegraphics[width=0.98\textwidth]{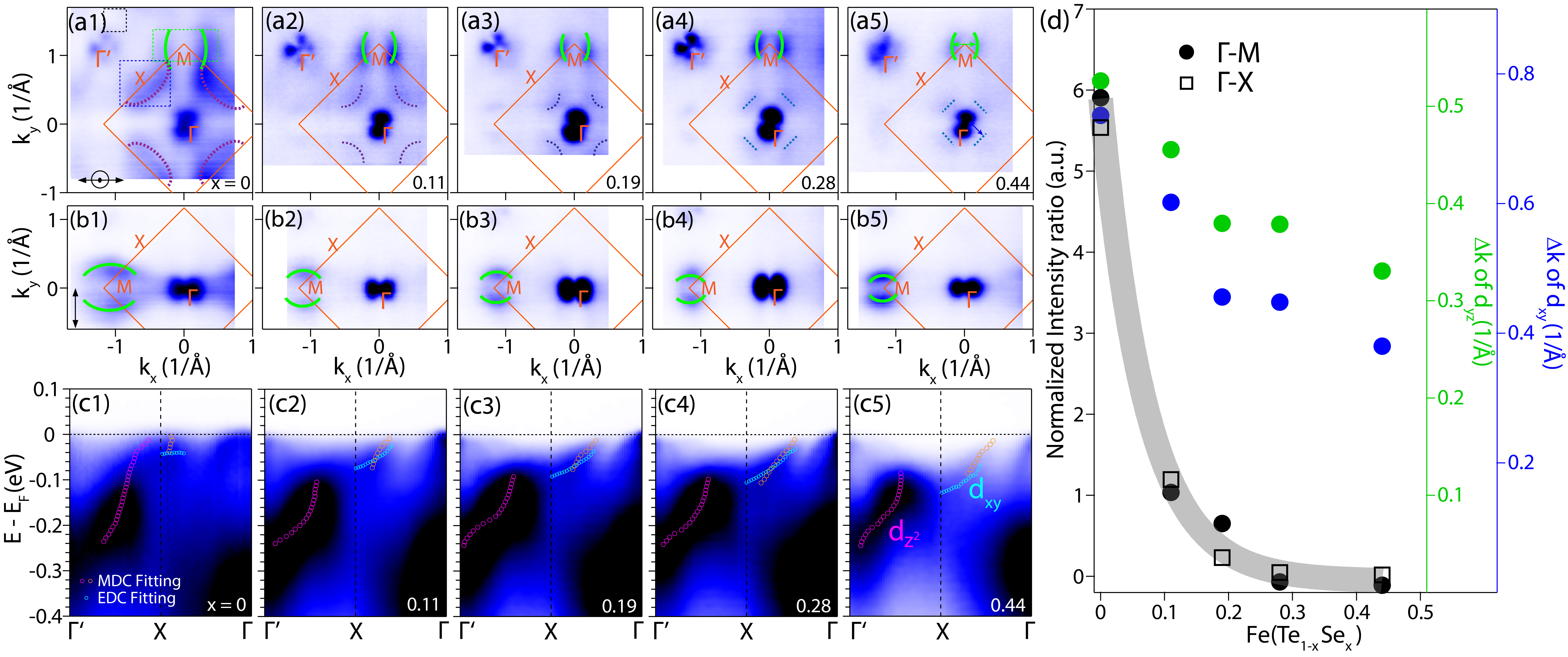}
\caption{\label{fig:Fig2}Fermi surface evolution and related spectral weight change with Se ratio $\textit{x}$. (a1)-(a5) Measured Fermi surfaces for different Se ratio \textit{x} under linear horizontal polarization with out-of-plane component with 72 eV photons. The orange squares mark the 2-Fe BZ. (b1)-(b5), Same as (a1)-(a5) except with linear vertical polarization. (c1)-(c5) Band image along $\Gamma$-X-$\Gamma$ overlaid with the band dispersions obtained by MDC and EDC fittings. (d) Fermi surface size and spectral weight change as a function of Se ratio \textit{x}. The solid dark circles represent the ratio between the area-averaged spectral weight within the blue dashed region at X and that of the green dashed region at M after subtracting an area-averaged background from the dark dashed region in (a). The empty dark squares are extracted in the same manner from an independent set of measured Fermi surfaces using a mixed polarization shown in SI Fig. S2~\cite{FTS_Supplement}. The green circles mark the extracted size of the $d_{yz}$ electron pocket as marked in (a5), while the blue circles plot the $k_F$ points of the band crossing along $\Gamma$-X as marked in (a5).}
\end{figure*}

FeTe$_{1-x}$Se$_x$ is a prototypical iron-based superconductor with the simplest crystal structure consisting of Fe-chalcogen layers~\cite{Hsu2008}. The Se-Te substitution is isoelectronic, retaining the $3d$-electron occupancy of $n$=6 per Fe ion. The parent compound, FeTe, is an antiferromagnetic (AFM) metal~\cite{Fruchart1975,Bao2009,Li2009,Liu2010}. Superconductivity emerges with the substitution of Se on Te sites, reaching a maximum $T_c$ of 14.5 K in FeTe$_{0.56}$Se$_{0.44}$~\cite{Liu2010,Martinelli2010}. With complete substitution of Se, FeSe is a superconductor below 8 K with a tetragonal to orthorhombic structural transition at 90 K~\cite{McQueen2009a} and no long-range magnetic order~\cite{McQueen2009}. As-grown single crystals of FeTe$_{1-x}$Se$_x$ especially for low values of \textit{x} have a tendency to harbor interstitial excess iron, which leads to spin-glass behavior and incoherence in the low energy electronic spectra. It has been shown that the excess Fe can be reduced or completely removed by either annealing in an oxygen or Te-vapor environment~\cite{Sun2015,Dong2011,Lin2015,Sun2015a}. The removal of excess Fe also suppresses the spin-glass region of the phase diagram, resulting in a phase diagram that bares a closer resemblance to that of iron pnictides. Here we adopt Te-vapor annealing method to treat all as-grown single crystals to reduce excess iron~\cite{Xu2018}. Previously, a temperature induced crossover to an OSMP in optimally-substituted FeTe$_{0.56}$Se$_{0.44}$ has been reported~\cite{Yi2015}. In this study, we utilize ARPES to study the low temperature behaviors of Fe$_y$Te$_{1-x}$Se$_x$ single crystals with varying Se content (nominally \textit{x} = 0, 0.11, 0.19, 0.28, 0.44, and 1). The Fe contents of the cleaved surface of these crystals are measured via energy dispersive x-ray spectroscopy following the ARPES measurements and determined to be \textit{y} = 1.08, 1.07, 1.03, 1.01, 1.00, and 1.00, respectively. For simplicity, we drop the \textit{y} index in further discussions.

\section{Methods}
The high quality FeTe$_{1-x}$Se$_x$ single crystal series was synthesized using the flux method~\cite{Liu2009}. The FeSe single crystal was synthesized using chemical vapor transport~\cite{Chen2019}. The excess Fe in as-grown FeTe$_{1-x}$Se$_x$ single crystals was reduced by annealing in Te vapor atmosphere~\cite{Koshika2013}. ARPES experiments were performed at beamline 5-4 of the Stanford Synchrotron Radiation Lightsource, beamline 4.0.3 and beamline 10.0.1 of the Advanced Light Source equipped with hemispherical electron analyzers. The angular resolution was set to 0.3$^\circ$. The total energy resolution was set to 15 meV or better. All samples were cleaved \textit{in-situ} at 10 K and all the measurements were conducted at ultra-high vacuum with a base pressure of 3 $\times$ 10$^{-11}$ torr.

\section{Results}
\subsection{Fermi Surface Evolution}

We begin by comparing and contrasting the measured FS of the optimally-substituted FeTe$_{0.56}$Se$_{0.44}$ with that of the paramagnetic state of FeTe (Fig.~\ref{fig:Fig1}). In FeTe$_{0.56}$Se$_{0.44}$, as previously reported, the FS consists of small hole pockets near the Brillouin zone (BZ) center ($\Gamma$) and electron pockets near the BZ corner (M). The wavevector difference between them matches the momentum transfer, where neutron spin resonance has been observed in the superconducting state~\cite{Qiu2009}. In FeTe, the apparent fermiology has drastically changed. While the hole pockets near $\Gamma$ remain, the other dominant spectral weight appears along the BZ boundary centered at the X point. We note that such a change is not due to the emergence of the magnetic order in FeTe, as the FS is measured above the ordering temperature, and distinct reconstructions are observed across the magnetic transition (see supplementary information (SI) Fig. S1). For the rest of the paper, we only discuss measurements on FeTe above the magnetic phase.

\begin{figure*}
\includegraphics[width=0.98\textwidth]{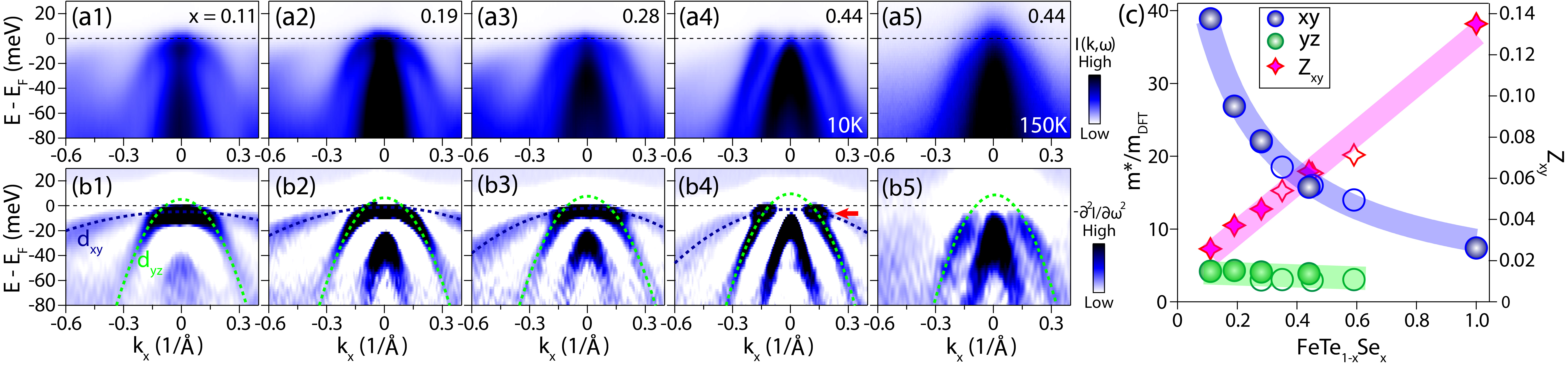}
\caption{\label{fig:Fig3}Orbital-dependent evolution of band effective mass in FeTe$_{x-1}$Se$_x$. (a1)-(a4) Band dispersions measured with 22 eV photons at 10 K along the $\Gamma$-M direction around the BZ center for different \textit{x}. (a5) Same measurement as (a4) except  taken at 150 K. (b1)-(b5) Corresponding second energy derivative images of (a1)-(a5). Parabolic fittings of the $d_{yz}$ and $d_{xy}$ hole bands are overlaid. The red arrow in (b4) points at the hybridization of $d_{yz}$ and $d_{xy}$ orbitals. (c) Extracted band enhancement factor $m^*$/m$_{DFT}$ for $d_{yz}$ and $d_{xy}$ are plotted as a function of \textit{x} represented by circles. The shaded lines are guides to the eyes. Diamond markers show the inverse of the $d_{xy}$ mass enhancement where the shaded pink line is a linear fit. The empty markers are reproduced from Ref. ~\cite{Liu2015a} }
\end{figure*}

To understand the evolution of the change shown in Fig.~\ref{fig:Fig1}, we present the measured FS of FeTe$_{1-x}$Se$_x$ with different Se concentrations across the phase diagram (Fig.~\ref{fig:Fig2}). The measurements under two distinct light polarizations reveal features with different orbital symmetries due to the photoemission matrix elements. Two qualitative changes can be observed with the decrease of the Se ratio (\textit{x}): i) expansion of the electron pockets near the M point, and ii) emergence of a broad arc-like feature near the X point. We quantify each in turn. The expansion of the electron pocket around M with decreasing Se ratio \textit{x} is evident in both sets of measurement under different polarizations. This expansion can be quantified by extracting the opening of the Fermi pocket in momentum (plotted as green solid circles in Fig.~\ref{fig:Fig2}d). The second evident FS evolution is the emergence of a broad arc-like feature near the X point of the BZ, marked by a dashed purple line in FeTe (Fig.~\ref{fig:Fig2}a1). From a consideration of photoemission matrix element effects from maps taken under three different polarizations, we attribute this feature to the d$_{z^2}$ orbital (see SI~\cite{FTS_Supplement}). Comparing the FS across different Se content level, this feature evolves from a faint intensity near $\Gamma$ that is the spectral weight from the outer $d_{xy}$ hole band for \textit{x} = 0.44. This is evident from the band dispersion measured along the $\Gamma$-X direction as shown in Fig.~\ref{fig:Fig2}c, where the Fermi momentum crossing ($k_F$) of a hole-like band from the $\Gamma$ point shifts toward the X point with decreasing \textit{x}. We employ two methods to quantify the evolution of this feature with Se content by tracking both the location of this feature in momentum and its spectral weight. To track the location, we extract the $k_F$ of this feature along the $\Gamma$-X direction as indicated by the blue arrow in Fig.~\ref{fig:Fig2}a5, and plot in Fig.~\ref{fig:Fig2}d. This distance increases with decreasing \textit{x} which confirms the systematic change. To quantify the emergence of the d$_{z^2}$ spectral weight near the X point, we take the ratio of the intensity around the X point and the intensity around the M point in the following way. We first obtain the integrated spectral weight normalized by area within the blue box around the X point and the green box around the M point (Fig.~\ref{fig:Fig2}a). From these we subtract a background from a featureless region in momentum space indicated by the black dashed box (Fig.~\ref{fig:Fig2}a). For better comparison across samples, we take the ratio of the background-subtracted spectral weight around X to that of the $d_{yz}$ band around M for each sample and plot the trend in Fig.~\ref{fig:Fig2}d. As Se content decreases, the intensity at X increases relative to that around M. The trend extracted using the same method for a set of FS measured under a rotated polarization shows the same behavior (SI Fig. S2~\cite{FTS_Supplement}). Taking all the trends extracted from the FS evolution (Fig.~\ref{fig:Fig2}d), a gradual rise is observed with decreasing Se ratio, and a more rapid change occurs below \textit{x} $\le$ 0.2. Such type of FS evolution contrasts with that of any other iron-based superconductor phase diagrams with symmetry-breaking phase transitions. To understand the origin of this evolution, we examine the orbital-dependent correlation effects.

\begin{figure*}
\includegraphics[width=0.8\textwidth]{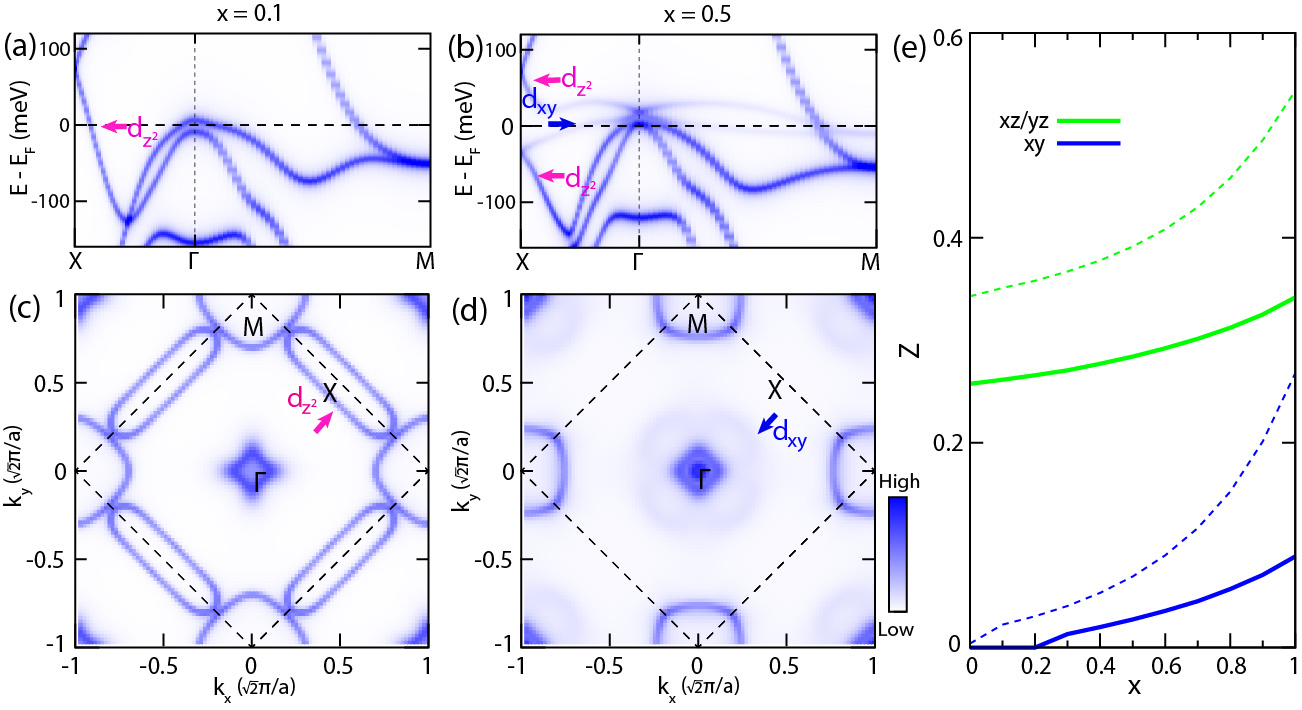}
\caption{\label{fig:Fig4}Theoretical calculations of FeTe$_{x-1}$Se$_x$. (a) Calculated band structure for the OSMP FeTe$_{0.9}$Se$_{0.1}$. The pink arrow indicates a dominantly $d_{z^2}$ band along $\Gamma$-X. (b) Same as (a) but for the strongly correlated metallic phase in FeTe$_{0.5}$Se$_{0.5}$, where the blue arrow shows the $d_{xy}$ band. (c) Calculated FS for FeTe$_{0.9}$Se$_{0.1}$, where reconstructed spectral weight is marked by a pink arrow. (d) Same as (c) but for FeTe$_{0.5}$Se$_{0.5}$. (e) Orbital-resolved coherence factor Z as a function of \textit{x}. Solid line represents U = 3 eV, dashed line represents U = 2.65 eV.}
\end{figure*}

\subsection{Orbital-Dependent Mass Enhancement}
We first examine the low-temperature (10 K) band renormalization across the phase diagram. Near the BZ center along the high symmetry direction $\Gamma$-M, three hole-like bands are observed near the Fermi level ($E_F$) for all Se content (Fig.~\ref{fig:Fig3}a). This can be better visualized from the second energy derivative plots (Fig.~\ref{fig:Fig3}b). Consistent with previous results~\cite{Nakayama2010}, these three bands from the innermost band to the outermost are identified as dominantly $d_{xz}$, $d_{yz}$, and $d_{xy}$, respectively. The dashed lines overlaid on the images are fittings to a parabolic curve of the band dispersion, with green representing the $d_{yz}$ orbital and blue the $d_{xy}$ orbital. As the phase diagram is traversed along the substitution-axis at 10 K, it is evident that the band curvature of the $d_{yz}$ band does not vary strongly while that of the $d_{xy}$ band flattens considerably with decreasing Se content. Since the effective mass for each band is proportional to the inverse of the band curvature, we can extract the orbital-dependent mass enhancement by taking the ratio between the fitted band effective mass from experimental data $m^*$ and the corresponding band effective mass from first-principle calculations $m_{DFT}$. The resulting orbital-dependent mass enhancement as a function of Se content is plotted in Fig.~\ref{fig:Fig3}c, which is consistent with the trend from previously reported results for higher Se content~\cite{Maletz2014,Liu2015a}. First, the mass enhancement of $40$ is unusually large for 3$d$-electrons near the occupancy $n=6$, even for the iron chalcogenides. Second, we observe an orbital-dependent band renormalization. The $d_{xy}$ orbital has a much larger band renormalization factor than the $d_{yz}$ orbital. Third, while the mass enhancement of the $d_{yz}$ orbital rises slowly with decreasing Se ratio \textit{x}, a divergent behavior of the mass enhancement for the $d_{xy}$ orbital is observed with decreasing Se content. We also plot the inverse of the $d_{xy}$ mass enhancement, showing a linear trend versus Se ratio with a fitted intercept of \textit{x} = -0.1. These results strongly indicate that FeTe is in proximity to an OSMP ground state and suggest that the evolution of the apparent fermiology is associated with the gradual disappearance of the $d_{xy}$ spectral weight as \textit{x} is decreased.

In addition, a previous report shows that at \textit{x} = 0.44, the $d_{xy}$ spectral weight gradually disappears with increasing temperature~\cite{Yi2015}. Indeed, when measured at 150 K, the spectral weight for the $d_{xy}$ hole band disappears, leaving the itinerant $d_{xz}$ and $d_{yz}$ hole bands near $\Gamma$ (Fig.~\ref{fig:Fig3}b5), suggestive of a temperature induced crossover into the OSMP. A similar disappearance of the $d_{xy}$ spectral weight is also observed for \textit{x} = 0.25, with a lower crossover temperature than that of \textit{x} = 0.44 (See SI Fig. S3~\cite{FTS_Supplement}), suggesting that the ground state of \textit{x} = 0.11 is closer to that of the OSMP than that of the \textit{x} = 0.44 compound.

\subsection{Visualizing the Selective Localization of $d_{xy}$-Electrons via $d_{z^2}$-Electrons}
The orbital-dependent band renormalization, nearly divergent $d_{xy}$ band effective mass, and redistribution of spectral weight around the FS, taken together can be understood as manifestations of the tendency of the FeTe$_{1-x}$Se$_x$ system towards an OSMP when Se is gradually replaced with Te. In the presence of relatively strong electron correlations, orbital-dependent correlations become strong, where the $d_{xy}$ orbital-dominated band near $E_F$ gradually diminishes together with the divergence of its mass enhancement. This disappearance of the $d_{xy}$ states near $E_F$ in turn results in a redistribution of the residual electronic states at $E_F$ as the FeTe end of the phase diagram is approached. But why should the diminishing $d_{xy}$ weight at the Fermi level be seen through the crossing of the $d_{z^2}$ band through the Fermi level? Qualitatively, we attribute this to the hybridization picture. Recall that symmetry dictates a nonzero hybridization matrix between the $d_{xy}$ and $d_{z^2}$ states near $X$. Just like the $f$-electron states in a heavy fermion metal produces a "large" FS, the itinerant $d_{xy}$ states hybridize the $d_{z^2}$ state near $X$ in the Fermi-surface formation. However,  just like the localization of the $f$-electrons in heavy fermion systems leads to a "small" FS of the $spd$ conduction electrons, localizing the $d_{xy}$-electron state allows the $d_{z^2}$-electron state to cross the Fermi level near $X$.

To substantiate this qualitative picture, we have carried out theoretical calculations based on a five-orbital Hubbard model for FeTe$_{1-x}$Se$_x$ (Section VI of SI~\cite{FTS_Supplement}). We find that the substitution of larger Te atoms for smaller Se atoms increases the Fe-Se/Te bond length and decreases the Fe-Te/Se-Fe bond angle. The former effect increases the overall correlation strength by lengthening the dominant hopping path while the latter effect suppresses the effective hopping for the largely in-plane $d_{xy}$ orbital more than that of the other orbitals, pushing the system toward an OSMP. 

All salient features of our experimental observations presented are captured in this set of calculations, as shown by a direct comparison of the calculations for the correlated metallic phase and the OSMP (Fig.~\ref{fig:Fig4}). In the strongly correlated metallic phase calculated for \textit{x} = 0.5 (Fig.~\ref{fig:Fig4}b), the $d_{xy}$ dominated bands are strongly renormalized (blue arrow). The calculations at a $U$ of $~$3 eV are consistent with experimentally observed band renormalizations (Section VI of SI~\cite{FTS_Supplement}). When projected onto a FS mapping within a finite integration energy window about $E_F$, the flattened $d_{xy}$ hole pocket intensity becomes largely incoherent and enlarged due to its strongly renormalized bandwidth, resulting in a large faint hole pocket around the $\Gamma$ point (blue arrow in Fig.~\ref{fig:Fig4}d). When the OSMP phase is entered, the $d_{xy}$ band intensity disappears from $E_F$ (Fig.~\ref{fig:Fig4}a). The hybridization between the heavy $d_{xy}$ band and other orbitals disappears, leading to the appearance of restructured FS. This is most apparent along the $\Gamma$-X direction. In the strongly correlated metallic phase (Fig.~\ref{fig:Fig4}d), the strongly renormalized $d_{xy}$ cross $E_F$ along $\Gamma$-X. In the OSMP (Fig.~\ref{fig:Fig4}c), a band from higher binding energy of dominantly d$_{z^2}$ orbital indeed rises to $E_F$, forming the large pocket around the X point, consistent with our observed emergence of the arc-like feature around the X point. Regarding the concomitant evolution of the band dispersions, in the regime approaching the OSMP, the $d_{xy}$ bandwidth would narrow while its spectral weight diminishes, causing the hybridization between this $d_{xy}$ band and the highly dispersive d$_{z^2}$ band to decrease. This is consistent with our observed dispersions along the $\Gamma$-X direction (Fig.~\ref{fig:Fig2}c and SI Fig. S4~\cite{FTS_Supplement}). For \textit{x} = 0.44, both momentum distribution curve (MDC) and energy distribution curve (EDC) fitting of the hole-like dispersion near $\Gamma$ agree well. With decreasing \textit{x}, the fitting between MDCs and EDCs becomes increasingly distinct. As MDC fitting identifies better the steep dispersions while EDC flatter dispersions~\cite{Zhang2008}, the increasing disagreement is consistent with the tendency of an increasingly flat $d_{xy}$ band and the rising up of a dispersive d$_{z^2}$ band. Simultaneously, the calculated $d_{xz}$/$d_{yz}$ electron pocket enlarges, as observed experimentally. As a function of \textit{x}, the calculated coherence factor, Z, for the $d_{xz}$/$d_{yz}$ orbital is compared to that of the $d_{xy}$ orbital (Fig.~\ref{fig:Fig4}e), where localization of the $d_{xy}$ orbital appears for \textit{x} $<$ 0.2 with U $\sim$ 3 eV.

\begin{figure}
\includegraphics[width=0.48\textwidth]{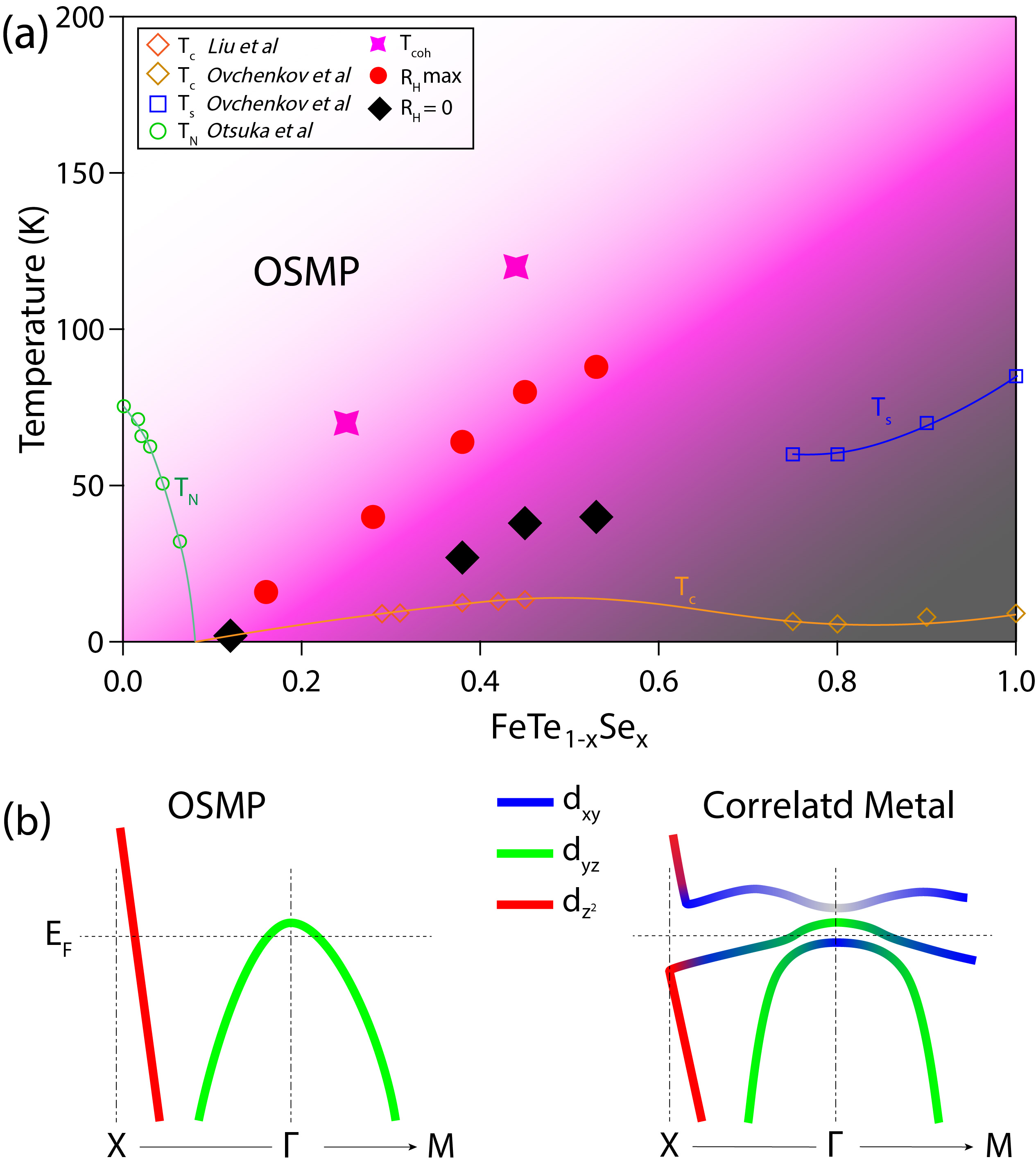}
\caption{\label{fig:Fig5}Phase diagram of FeTe$_{x-1}$Se$_x$. (a) $T_N$~\cite{Otsuka2019}, $T_s$~\cite{Ovchenkov2019} and $T_c$~\cite{Liu2010,Ovchenkov2019} represent the bi-collinear spin-density wave transition, tetragonal to orthorhombic structure transition, and superconducting transition adapted from previous reports. The temperatures for $R_H$ max and $R_H$ = 0 are extracted from the Hall resistivity measurements~\cite{Jiang2020}. $T_{coh}$ is the temperature at which the photoemission spectral weight of $d_{xy}$ orbital is observed to vanish (SI Fig. S3~\cite{FTS_Supplement}). The background with gradient color suggests the OSMP crossover. (b) Schematics showing the key features of the band structure of FeTe$_{x-1}$Se$_x$ in the OSMP and correlated metal phase respectively. The disappearance of band hybridization due to the complete incoherence of d$_{xy}$ orbital is responsible for the FS reconstruction. }
\end{figure}

We stress that, the selective localization of the $d_{xy}$-electron states also causes the suppression of their hybridization with $d_{xz}$/$d_{yz}$-electron states. Indeed, a hybridization between the $d_{xy}$ and $d_{yz}$ hole bands is observed near \ef~in FeTe$_{1-x}$Se$_x$ with \textit{x} = 0.44 (Fig.~\ref{fig:Fig3}b4) and diminishes towards FeTe. The $d_{xz}$/$d_{yz}$-electron states are less dispersive compared with $d_{z^2}$ and the hybridization occurs near the band top and is thus less discernable. Moreover, the $d_{xz}$/$d_{yz}$ bands already cross $E_F$ and form a hole pocket around $\Gamma$ in the presence of hybridization with the $d_{xy}$ band (Fig.~\ref{fig:Fig3}b). The suppression of the hybridization due to the localization of the $d_{xy}$ orbital in the OSMP does not therefore lead to a drastic FS reconstruction but rather the disappearance of the $d_{xy}$-electron states near $E_F$.
However, because the $d_{z^2}$-electron states are considerably more dispersive and gapped-out from $E_F$ due to the hybridization with $d_{xy}$, they serve as an especially transparent diagnostic of the OSMP in the way of FS reconstruction (Fig.~\ref{fig:Fig5}b).

\section{Discussions and Summary}
Finally, taking all the results together, we arrive at the phase diagram for the FeTe$_{1-x}$Se$_x$ material family (Fig.~\ref{fig:Fig5}). While previous results~\cite{Yi2015} indicate that for optimally substituted compound at \textit{x} = 0.44, the OSMP can be reached via raised temperature at a characteristic temperature of 110 K, our observations reported here show that in the low temperature limit, the replacement of Se by Te also leads the material system towards the OSMP. This is supported by spectroscopic evidence of strongly orbital-selective band renormalization where the $d_{xy}$ effective mass tends toward a divergent behavior as the FeTe end is approached. Concomitantly, spectral weight from other orbitals redistribute near the $E_F$ to replace the diminishing $d_{xy}$ spectral weight. The characteristic crossover temperature for the OSMP therefore decreases with decreasing \textit{x}. This is confirmed by our measurement for a sample with \textit{x} = 0.25, where the temperature identified by the disappearance of $d_{xy}$ is measured to be 70 K (SI Fig. S3~\cite{FTS_Supplement}). 

Our understanding of the evolution from an orbital-dependent correlated metallic phase to an OSMP across the FeTe$_{1-x}$Se$_x$ phase diagram as observed from ARPES is also consistent with results reported by other probes. It has been reported that the compensated parent compounds of multi-band iron-based superconductors exhibit a negative Hall coefficient, $R_H$, due to the dominance of the electron mobility. This is observed in isovalent-substituted BaFe$_2$(As,P)$_2$, where across the entire phase diagram, $R_H$ remains negative~\cite{Fang2009,Kasahara2010}. For FeTe$_{1-x}$Se$_x$, however, there is a systematic change that varies as a function of substitution~\cite{Jiang2020}. This is clearly shown in the measurement over the phase diagram of Te-vapor treated FeTe$_{1-x}$Se$_x$~\cite{Jiang2020}. Contrary to the behavior in BaFe$_2$(As,P)$_2$, Hall resistivity measurements on FeTe$_{1-x}$Se$_x$ exhibit a maximum and a subsequent sign-change at lower temperatures while the total charge carrier remains neutral. This turn-over behavior marked by the maximum indicates the onset of a competing behavior. This characteristic temperature can be explained by the decoherence of the $d_{xy}$ portions of the electron pocket around the M point, leading to a reduction in the electron contribution to $R_H$. An alternative interpretation of the negative Hall coefficient in the iron pnictides is the ($\pi$, $\pi$) spin fluctuations that make the hole carriers behave like electron carriers in the Hall measurements~\cite{Fanfarillo2012}. The localization of the $d_{xy}$ orbital could change the fermiology away from the nesting condition associated with the ($\pi$, $\pi$) spin fluctuations. This in turn suppresses the ($\pi$, $\pi$) spin fluctuations, which reduces the negative contribution to the Hall coefficient. This characteristic temperature scale extracted from $R_H$ exhibits a similar trend with Se ratio \textit{x}, similar to $T_{coh}$, which is the temperature where the spectral weight of $d_{xy}$ orbital vanishes with increasing temperature. 

We also note that in a recent magneto-transport measurement, such coherent-incoherent crossover is also reported~\cite{Otsuka2019}. It has also been pointed out that this coherent-incoherent crossover temperature scale of the $d_{xy}$ orbital is coupled to the $B_{2g}$ nematic susceptibility measured by elastoresistance, and hence suggests the potential important role played by the $d_{xy}$ orbital to nematicity in the iron-based superconductors~\cite{Jiang2020}. 

Previous studies of both Dynamic Mean Field Theory (DMFT) and ARPES have also identified the incoherent spectral weight in the high binding energy regime of FeSe as the lower Hubbard band (LHB), indicating the presence of strong electron correlations in this material system~\cite{Evtushinsky2016,Watson2017}. The incoherent feature has also been observed by our ARPES measurements as well as theoretical calculations (SI Fig. S10~\cite{FTS_Supplement}).

Further connection between our ARPES report of OSMP and magnetic excitations via inelastic neutron scattering can also be made. It has been reported that the total magnetic spectral weight at 300 K in FeTe is reduced by roughly half at 10 K, signaling a change in the number of localized electrons as in the scenario of an OSMP~\cite{Zaliznyak2011,Tranquada2020}. Moreover, along both the temperature axis~\cite{Xu2014} and substitution~\cite{Christianson2013} axis, low energy magnetic excitations have been observed to change from a commensurate U-shape to a double-rod shape in the absence of any symmetry-breaking orders. The trend of such behavior is consistent with the observation of the modification of the low energy electronic states along both axes as we have reported, suggesting a common origin.

An OSMP is depicted by the coexistence of itinerant orbital and localized orbital within one material system~\cite{Anisimov2002,Mukherjee2016,Qiao2017}. In the FeTe$_{1-x}$Se$_x$ case, it is the itinerant $d_{xz}$/$d_{yz}$ orbitals and the localized $d_{xy}$ orbital. In our work, we have also discovered a surprisingly transparent way for the $d_{z^2}$ electrons to serve as a diagnostic for the selective localization of the $d_{xy}$ electron states as the OSMP emerges (Fig.~\ref{fig:Fig5}b). To achieve an OSMP in FeTe$_{1-x}$Se$_x$, besides the strong electron-electron interaction, three other ingredients are necessary: i) Crystal field splitting that separates the $d_{xy}$ orbital from $d_{xz}$/$d_{yz}$; ii) Bare bandwidth of the $d_{xy}$ orbital being narrower than that of the $d_{xz}$/$d_{yz}$ orbitals; iii) Suppression of the inter-orbital interactions due to Hund’s J~\cite{Si2016}. The observation of a low temperature tunable pathway towards an OSMP in FeTe$_{1-x}$Se$_x$ is somewhat reminiscent of the tunability across a quantum critical point in the heavy fermion systems where the heavy \textit{f}-electrons coexist with the itinerant \textit{d}-orbital electrons and the hybridization between the two depends on a non-thermal tuning parameter. In the vicinity of such a quantum critical point, a variety of magnetic and superconducting phases are often found to emerge. The observation of the proximity to such an OSMP ground state in the iron-chalcogenide system may likewise provide the basis to understand its exotic emergent phases.

\section{acknowledgments}
We are thankful to Yu He for enlightening discussions. ARPES experiments were performed at the Advanced Light Source and the Stanford Synchrotron Radiation Lightsource, which are both operated by the Office of Basic Energy Sciences, U.S. DOE. Work at University of California, Berkeley and Lawrence Berkeley National Laboratory was funded by the U.S. Department of Energy, Office of Science, Office of Basic Energy Sciences, Materials Sciences and Engineering Division under Contract No. DE-AC02-05-CH11231 within the Quantum Materials Program (KC2202) and the Office of Basic Energy Sciences. The ARPES work at Rice University was supported by the Robert A. Welch Foundation Grant No. C-2024 as well as the Alfred P. Sloan Foundation. The materials synthesis efforts at Rice are supported by the US Department of Energy (DOE), Basic Energy Sciences (BES), under Contract No. DE-SC0012311 and the Robert A. Welch Foundation, Grant No. C-1839 (P.D.). Theory work at Rice University is supported by the U.S. Department of Energy, Office of Science, Basic Energy Sciences, under Award No. DE-SC0018197, and by the Robert A. Welch Foundation Grant No. C-1411. Theory work at Renmin University has in part been supported by the National Science Foundation of China Grant No. 11674392, Ministry of Science and Technology of China, National Program on Key Research Project Grant No.2016YFA0300504 and Research Funds of Renmin University of China Grant No. 18XNLG24 (R.Y.). Theory work at Los Alamos was carried out under the auspices of the U.S. Department of Energy National Nuclear Security Administration under Contract No. 89233218CNA000001, and was supported by the LANL LDRD Program. Work at the University of Washington was supported by NSF MRSEC at UW (DMR1719797). Work at Sun Yat-Sen University was supported by the National Natural Science Foundation of China (Grant No. 11904414), Natural Science Foundation of Guangdong (No. 2018A030313055), National Key Research and Development Program of China (No. 2019YFA0705700), and Young Zhujiang Scholar program. Work at Brookhaven is supported by the Ofﬁce of Basic Energy Sciences, Division of Materials Sciences and Engineering, U.S. Department of Energy under contract Nos. DE-AC02-98CH10886 and DE-SC00112704.


\end{document}